\documentclass[fleqn,usenatbib]{mnras}

\usepackage{newtxtext,newtxmath}

\usepackage[T1]{fontenc}

\DeclareRobustCommand{\VAN}[3]{#2}
\let\VANthebibliography\thebibliography
\def\thebibliography{\DeclareRobustCommand{\VAN}[3]{##3}\VANthebibliography}

\newcommand{\bdv}[1]{\mbox{\boldmath$#1$}}

\usepackage{graphicx}	
\usepackage{amsmath}	
\usepackage{hyperref}
\usepackage{caption}
\usepackage{subcaption}
\usepackage{lipsum}
\usepackage[table]{xcolor}
\usepackage{adjustbox}
\usepackage{rotating}
\usepackage[online]{threeparttable}
\usepackage{CJK}

\title[Binary BHs via microlensing]{Identification of stellar-mass black hole binaries and the validity of linear orbital motion approximation in microlensing}

\author[Ma, Zhu, \& Yang]{
Xiaoyi Ma (马潇依)$^{1,2}$,
Wei Zhu (祝伟)$^{1,2}$\thanks{Email: weizhu@tsinghua.edu.cn},
and
Hongjing Yang (杨弘靖)$^1$
\\
$^{1}$ Department of Astronomy, Tsinghua University, Beijing 100084, China\\
$^{2}$ Canadian Institute for Theoretical Astrophysics, University of Toronto, Toronto,
Ontario M5S 3H8, Canada
}

\date{Accepted XXX. Received YYY; in original form ZZZ}

\pubyear{2022}

\begin{document}
\begin{CJK*}{UTF8}{gbsn}
\label{firstpage}
\pagerange{\pageref{firstpage}--\pageref{lastpage}}
\maketitle

\begin{abstract}
Gravitational microlensing is unique in detecting binary black (BH) holes with wide (a few au) separations. Models predict that about 1\% of microlensing binaries should be due to binary BHs, and yet zero has been robustly identified. Using simulated events with binary BH lenses, we show that the microlensing parallax effect in a typical binary BH event cannot be reliably detected. Given the crucial role of the parallax parameter in determining the mass of dark microlenses, this may explain the non-detection of binary BHs. Additionally, we show that in only a small fraction ($\lesssim7\%$) of the simulated events the full orbital motion of the binary lens cannot be modeled with the linear orbital motion approximation. This approximation has been frequently used in modelings of binary microlensing events. 
\end{abstract}




\section{Introduction}\label{section1}

Black holes (BHs) are generally considered the end product of the evolution of massive stars, as such, the discovery of such stellar-mass BHs can therefore help constrain the massive star evolution \citep[e.g.,][]{Abbott_2016,Thompson:2019,MillerJones:2021}.
So far there have been over 100 detections of stellar-mass black holes, mostly from observations of X-ray binaries and the LIGO/Virgo detections of the gravitational wave radiations \citep[e.g.,][]{blackcat:2016,Abbott_2019,Abbott_2021}.

Not relying on the (electromagnetic or gravitational) radiation from the target object/system, gravitational microlensing has its unique advantage in detecting extremely faint and even dark objects \citep{Paczynski:1986,Mao:1991}. In terms of stellar-mass BHs, microlensing is arguably the only technique that can detect isolated BHs as well as BHs in wide (a few au) binaries with dark companions. Such objects are expected to be over-represented in microlensing surveys because their massive nature leads to larger cross sections and longer event durations. Indeed, it is estimated that about 1\% of all microlensing events are produced by BH lenses \citep{Gould:2000}.

Thanks to the wide-field photometric surveys starting from the early 1990s, over 20,000 microlensing events have been discovered toward the direction of the Galactic bulge, of which $\sim10\%$ may show signatures of binary microlensing \citep[e.g.,][]{Sumi:2011, Udalski:2015, Kim:2016}. According to \citet{Gould:2000}, these numbers would suggest the detection of nearly 200 isolated stellar-mass BHs and about 20 binary BHs with wide separations.
\footnote{It is unclear whether the fraction of BHs in wide binaries is the same as the fraction of stars in wide binaries. On the one hand, massive stars that end up in BHs are predominantly in binaries \citep[e.g.,][]{DucheneKraus:2013}. On the other hand, binary stellar systems that are loosely bound may be easily disrupted on their way to form binary BHs \citep[e.g.,][]{Olejak:2020}.}
However, except for about one dozen candidates \citep[e.g.,][]{Agol:2002, Bennett:2002, Mao:2002, Poindexter:2005, Dong:2007, Shvartzvald:2015, Wyrzykowski:2016, Mroz:2021}, no robust detections of stellar-mass BHs has been made via microlensing.
\footnote{During the review process of this paper, \citet{Sahu:2022} reported that MOA-2011-BLG-191/OGLE-2011-BLG-0462 was produced by an isolated stellar-mass black hole, based on astrometric microlensing from Hubble Space Telescope and photometric microlensing from the ground. An independent analysis by \citet{Lam:2022} claimed that the lens mass might be lower and that a neutron star could not be ruled out.}

In order to determine the mass of a microlens (or a lens system), $M_{\rm L}$, and thus identify it as a BH object, one needs to measure (or constrain) two microlensing parameters, the angular Einstein radius $\theta_{\rm E}$ and the microlensing parallax $\pi_{\rm E}$ \citep{Gould:1992,Gould:2000b,Gould:2004}
\begin{equation}\label{eq1}
    M_{\rm L} = \frac{\theta_{\rm E}}{\kappa \pi_{\rm E}} ;\quad
    \kappa \equiv \frac{4G}{c^2 \rm au} \approx 8.14 \frac{\rm mas}{M_\odot}
\end{equation}
Here
\begin{equation}\label{eq2}
    \theta_{\rm E} \equiv \sqrt{\kappa M_{\rm L} \pi_{\rm rel}};\quad
    \pi_{\rm E} \equiv \frac{\pi_{\rm rel}}{\theta_{\rm E}},
\end{equation}
where $\pi_{\rm rel} \equiv {\rm au}(D_{\rm L}^{-1}-D_{\rm S}^{-1})$ is the relative parallax between the lens and the source, and $D_{\rm L}$ and $D_{\rm S}$ are the distances to the lens and the source, respectively. This work focuses on the determination of the microlensing parallax $\pi_{\rm E}$, although measuring the angular Einstein radius $\theta_{\rm E}$ can also be challenging, especially for isolated microlenses.

One common way of measuring the microlensing parallax is to search for the imprints of the orbital motion of Earth around Sun on the microlensing light curve (\citealt{Gould:1992}; but see also \citealt{Refsdal:1966, Gould:1994}). This annual parallax method typically works better for microlensing events with longer durations, and thus is seemingly in favor of events involving BH lenses. However, the amplitude of the microlensing parallax is also reduced with the increasing lens mass, leading to the increased difficulty to reliably detect the parallax effect for truly BH events. Indeed, it has been shown that the microlensing parallax effect is usually undetectable in the case of isolated BHs \citep{Karolinski_2020}.

This work will check the detectability of the parallax effect in typical microlensing events produced by binary BHs.
Because we have nearly no knowledge of the wide-orbit BH binaries in the Milky Way from direct observations, we adopt results from the population synthesis simulation \citep[e.g.,][]{Lam:2020} for the distribution of wide-orbit BH binaries as well as the typical microlensing parameters they predict.
Similar to the isolated BH case, the amplitude of the microlensing parallax is small for events with binary BHs. The orbital motion effect of the binary lens system may further complicates the analysis. Given the very similarity between the annual parallax and lens orbital motion effects in modifying the microlensing light curves, it is possible that the signal from the lens orbital motion may be misinterpreted as the signal from the annual parallax, thus resulting in erroneous parallax parameters.

This paper is organized in the following way. We describe the details of our light curve simulation and binary modeling in Section~\ref{sec:method} and present the main results of this paper in Section~\ref{sec:result}. The implications of our results are discussed in Section~\ref{sec:discussion}.

\section{Methods} \label{sec:method}

\subsection{Light curve generation} \label{section2.1}

A standard binary-lens microlensing event is typically described by the following parameters
\begin{equation}
 \{ t_0,~u_0,~t_{\rm E},~\rho,~q,~s_0,~\alpha_0\}    
\end{equation}
in addition to the flux parameters $(F_{\rm S},~F_{\rm B})$. Here $t_0$ and $u_0$ are the time and distance at the closest approach between the source and the center of mass of the lens system, respectively, $t_{\rm E}$ is the Einstein ring crossing time, normalized to the total mass of the lens system, $\rho$ is the source size scaled to the Einstein ring radius, $q$ is the mass ratio between the binary component, $s_0$ is the projected separation between the two binary components scaled to the Einstein ring radius at $t_0$, and $\alpha_0$ is the angle between the source trajectory and the binary axis at $t_0$.
The direction of $\alpha_0$ follows the convention of \citet{Poleski:2019}, which is offset by $180^\circ$ from the convention of \citet{Skowron:2011}

The flux parameter $F_{\rm S}$ quantifies the flux of the source star that would be magnified during the event, and $F_{\rm B}$ is the flux from the other objects that contribute to the photometric aperture (i.e., companion to the source, companion to the lens, or ambient stars). When the microlensing parallax effect is considered, two additional parameters $(\pi_{\rm E,E},~\pi_{\rm E,N})$ are also included. They are the eastern and northern components of the parallax vector $\bdv{\pi_{\rm E}}$, respectively. The direction of parallax is the same as the relative lens-source proper motion.

In this work, we set $t_0$ to be ${\rm HJD} = 2459031.5$, which is July 1st, 2020 and thus approximately the middle of the annual microlensing season. We also choose typical values for the other parameters. Specifically, we set $u_0=0.3 ,~\rho=10^{-3},~t_{\rm E}=100\,{\rm days}, q=1$ and $s_0=0.8$. The event is assumed to occur at the equatorial coordinates $({\rm RA},~{\rm Dec})=(18^{\rm h}00^{\rm m}00^{\rm s},~-30^\circ 00^\prime 00^{\prime\prime})$, and the microlensing parallax parameters are $\pi_{\rm E,E}=\pi_{\rm E,N}=0.03/\sqrt{2}$. The amplitude of the microlensing parallax and the event timescale are chosen to be representative of a typical event by stellar-mass black holes (see Figure~13 of \citealt{Lam:2020}). For our chosen binary with $M_1=M_2=5\,M_\odot$, these values correspond to a lens distance of $D_{\rm L}=5\,$kpc (with the source in the bulge at $D_{\rm S}=8\,$kpc), an Einstein radius of $r_{\rm E}=12\,$AU, and a projected separation $r_{\perp,0}\equiv s_0 r_{\rm E}=9.6\,$AU at $t_0$.

We include the full orbital motion of the lens binary in the light curve generation. For simplicity, we fix the orbital eccentricity $e=0.3$, which is a typical value for stellar binaries \citep[e.g.,][]{DucheneKraus:2013}. We randomly draw an orbital inclination $I$ from a $\sin{I}$ distribution between 0 and $\pi/2$. We also randomly draw from a uniform distribution between $0$ and $2\pi$ the mean anomaly at $t_0$, $l_0$, the argument of periapsis, $\omega$, as well as the binary axis orientation relative to the source trajectory at $t_0$, $\alpha_0$. The eccentric anomaly at $t_0$, $\mathcal{E}_0$, is then determined from solving the Kepler's equation
\begin{equation}
    l_0 = \mathcal{E}_0 - e\sin{\mathcal{E}_0} ,
\end{equation}
and the true anomaly at $t_0$, $f_0$, from
\begin{equation}
    \cos{f_0} = \frac{\cos{\mathcal{E}_0}-e}{1-e \cos{\mathcal{E}_0}} 
\end{equation}
with the understanding that $\sin{f_0}$ shares the same sign as $\sin{\mathcal{E}_0}$. The semi-major axis $a$ can be derived from the following equation
\begin{equation}
     r_{\perp,0} = a (1-e\cos{\mathcal{E}_0})\sqrt{\cos^2{(f_0+\omega)}+\sin^2{(f_0+\omega)} \cos^2{I}} .
\end{equation}
Similar to the commonly used geometry in microlensing, we set the $x$-axis to be the direction of the binary axis, and thus the longitude of ascending node $\Omega$ can be determined by
\begin{equation}
    \tan{\Omega} = - \tan{(f_0+\omega)} \cos{I} 
\end{equation}
with $\Omega$ and $-(f_0+\omega)$ in the same quadrant. With these Keplerian parameters, we can then proceed and determine the projected separation, $s$, and the orientation of the source trajectory relative to the binary axis, $\alpha$, at any given epoch. 

The \texttt{VBBinaryLensing} package inside the \texttt{MulensModel} code is then used to compute the microlensing magnification $A(t)$ \citep{Bozza:2010,Bozza:2018,Poleski:2019}. We assume a typical source star with a baseline magnitude $m_0=18$ and no blending \footnote{Although zero blending was assumed in the light curve simulation, it was not forced in the light curve modeling. We used the \texttt{MuLensModel} module \citep{Poleski:2019} to derive the flux parameters. In other words, the degeneracy between the blending parameter and other microlensing parameters 
is taken into account in the modeling.}, and convert the magnification to magnitude
\begin{equation}
    m(t) = m_0 - 2.5\log_{10}A(t) .
\end{equation}
The uncertainty in magnitude is given by
\begin{equation}
    \sigma_m(t) = \sqrt{\sigma_{\rm sys}^2 + \sigma_0^2 10^{0.8(m(t)-m_0)}} ,
\end{equation}
where $\sigma_{\rm sys}=0.004$ is the systematic floor, $\sigma_0=0.04$ is the uncertainty at the reference magnitude $m_0$. The above noise curve matches the theoretical noise curve of an OGLE-IV-like survey \citep{Udalski:2015}. 
Note that we do not randomize the simulated data points around the model values with the uncertainties given above. Randomizing the simulated data points around the model values would complicate the modeling procedures and add additional random noises to the results, but no scientifically useful benefit. The same argument has been presented in the previous simulation work of \citet{Henderson:2014}.

The light curve is sampled at a cadence of once per day and within an 1800-day total duration, which is long enough to constrain the microlensing parallax.
Our simulated light curves do not contain seasonal gaps in the data span. A gap of 2--3 months per year is typical for observations taken from the ground. However, such gaps will not affect the characterization of light curves with 100\,days (or longer) timescales at a dramatic level, as long as the primary binary features are reasonably covered with observations.
\footnote{Events that failed this condition will hardly be considered for binary BH detections, given that there may exist alternative models.}
If any, the inclusion of seasonal gaps will further reduce the detectability of microlensing parallax effect.

\begin{figure*}
\includegraphics[width=0.9\textwidth]{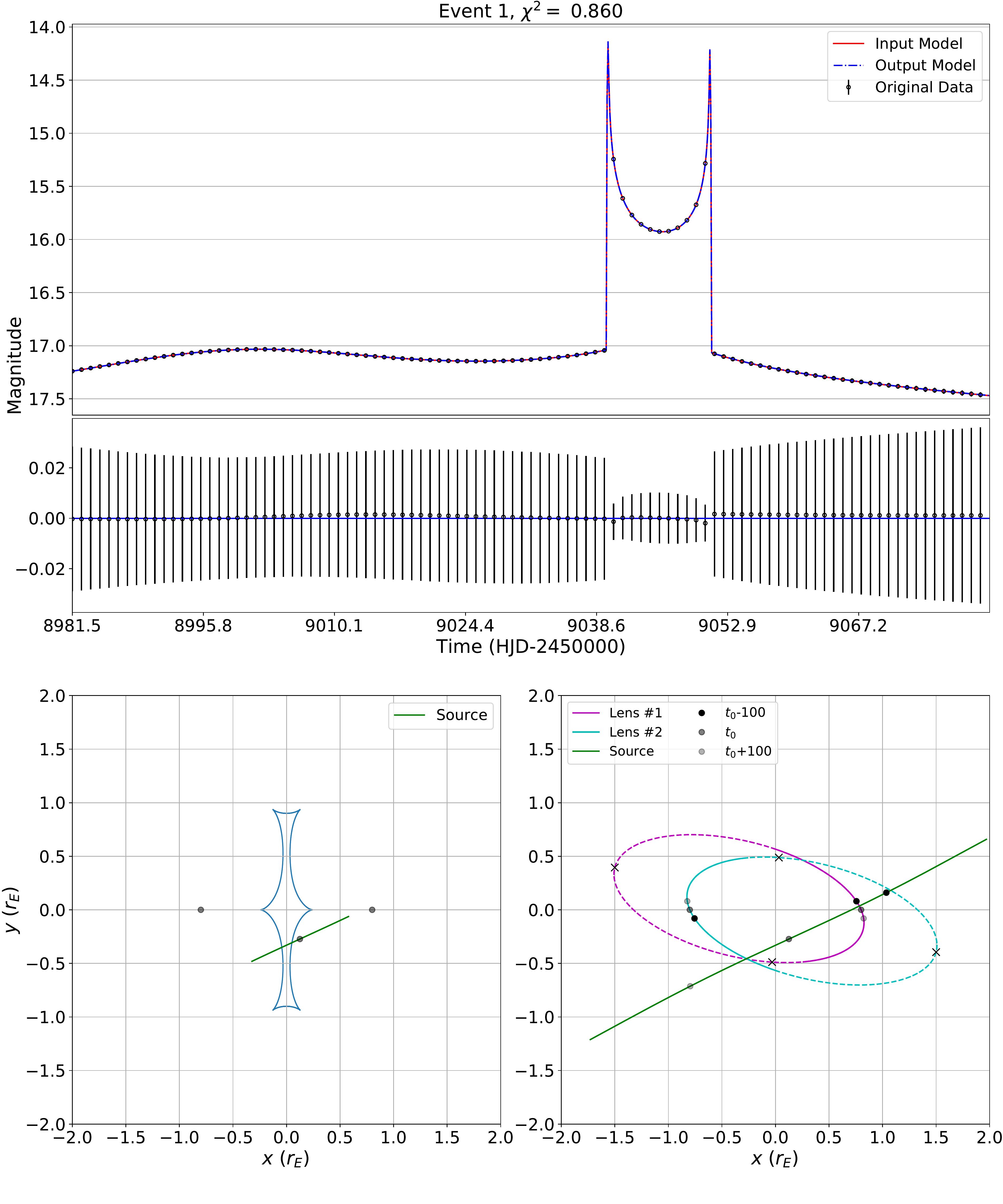}
\caption{The top panel shows the simulated light curve and the best-fit model of one example event. Data beyond the 100-day window centered on $t_0$ are not shown. The middle panel shows the residuals after the subtraction of the best-fit model. The bottom left panel is the lensing geometry at $t_0$. The source trajectory for the central 100-day window is shown in green, and the hexagonal-shaped caustic is shown in blue. The positions of the source and the two components of the binary lens at $t_0$ are shown as grey dots. The bottom right panel is similar to the bottom left panel, except that it shows the trajectories of both lenses during the full 1800-day window (solid line) and the trajectories of the source during 400-day window. The full orbits of both lenses are also indicated as dashed curves. The black crosses indicate the lens positions where the projected separation reaches the smallest (i.e., pericenter) and the largest (i.e., apocenter).}
\label{fig:example-lc}
\end{figure*}

In total 40 events are simulated, and Figure~\ref{fig:example-lc} illustrates one of them.
To check the robustness of our results, we also simulated 20 events with a different mass ratio $q=0.3$. The generation of this additional set of events follows closely the generation of the $q=1$ events.

\begin{figure*}
\includegraphics[width=0.8\textwidth]{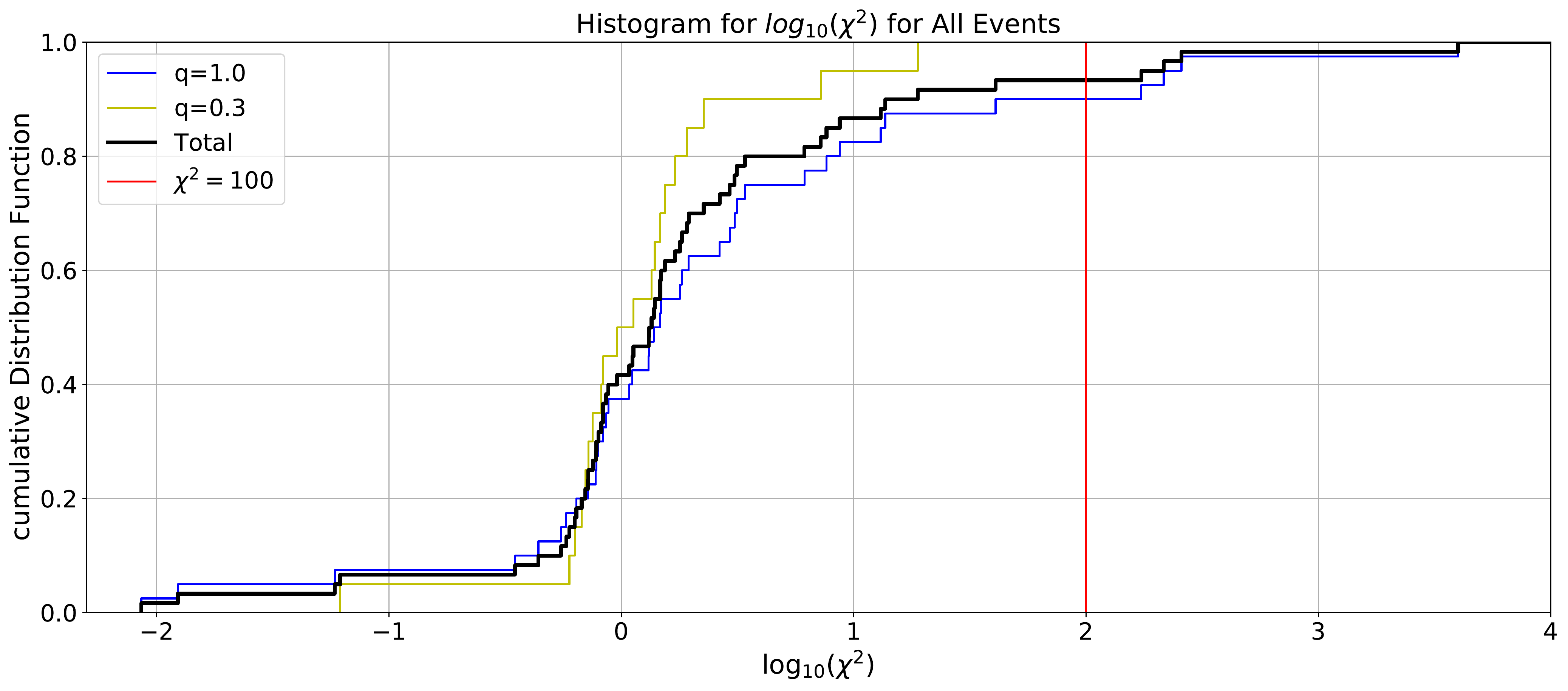}
\caption{The cumulative distribution of the best-fit model $\chi^2$ of all 60 simulated events. The red vertical line marks $\chi^2=100$, below which the associated linear orbital motion model is considered a reasonable approximation. The blue line indicates the events with $q=1$, the yellow line indicates the events with $q=0.3$, and the black line indicates the distribution for all events.}
\label{fig:chi2-cdf}
\end{figure*}

\subsection{Binary lens modeling} \label{section2.2}

The microlensing light curve arising from a binary lens is usually modeled with the following set of parameters
\begin{equation}
\{t_0,~u_0,~t_{\rm E},~\alpha_0\,~s
_0,~q,~\rho,~\pi_{\rm E,E},~\pi_{\rm E,N},~{\rm d}s/{\rm d}t,~{\rm d}\alpha/{\rm d}t\}
\end{equation}
The first seven of them are necessary to describe a static binary with no parallax effect, and the last four are introduced for the inclusion of microlensing parallax and lens orbital motion effects. Note that the orbital motion of the lens system is approximated with two linear parameters, ${\rm d}s/{\rm d}t$ and ${\rm d}\alpha/{\rm d}t$ \citep[e.g.,][]{Dong:2009,Skowron:2011}. This linear approximation seems to work except for some rare cases \citep[e.g.,][]{Skowron:2011,Shin:2011,Shin:2012,Han:2016}, primarily because the duration of the binary signature is much shorter than the orbital period of the lens binary. We will investigate how often this linear approximation breaks down in Section~\ref{sec:result}.

Following the standard approach in microlensing, we first fit for the binary parameters without higher-order effects and then introduce parallax and orbital motion parameters in the modeling. Considering that the cadence in our simulated events is long relative to the source crossing time, we fix the scaled source size $\rho$ to the input value. \footnote{We have run the MCMC modeling with free $\rho$ for the event with large $\chi^2$ and found no significant improvement in the goodness-of-fit, thus validating our decision on fixing $\rho$.} Our light curve modeling proceeds in the following way. First, we perform a grid search in the $(q,~s_0)$ plane to identify the set of parameters that yield the lowest $\chi^2$. This search is done for five values of $q$ and $s_0$ equally spaced in the range 0.9--1.1 and 0.75--0.85, respectively. At each grid point, we use the Nelder-Mead algorithm of \texttt{scipy.optimize} with no bounds \citep{Virtanen:2020} to search for the best set of $\{t_0,~u_0,~t_{\rm E},~\alpha_0\}$ that yields the lowest $\chi^2$ value. The set of parameters found from this grid search is then polished and all parameters but $\rho$ are set free to identify the best-fit global solution. Then, we perform Markov chain Monte Carlo (MCMC) analysis to determine the best-fit parameters and their associated uncertainties. The set of parameters from the previous step is used as the initial guess of this MCMC process, and the MCMC is done through the \texttt{emcee} package developed by \citet{ForemanMackey:2013}. For the majority of events we use 20 walkers, each with 5,000 samples. In cases of unsatisfactory MCMC results, such as unconverged chains and/or overly large $\chi^2$ values ($\chi^2>100$), we repeat the above procedures with finer ($11 \times 11$) grid and longer (40 walkers, each with 5,000 samples) chains.  

The first 50,000 elements of the flattened MCMC chain are treated as the burn-in steps, and we use the remaining chain elements to derive the best-fit values and the associated uncertainties.

After the above procedures, all but seven events have reasonably good models found with $\chi^2 \lesssim 100$. 
For the remaining seven, grid searches on a much larger parameter space are performed to identify better solutions (if any). Specifically, we search for local $\chi^2$ minima on the grid of $(\log{q}, \log{s}, \alpha_0)$, with $(21, 41, 20)$ values equally spaced in ranges of $-2 \leq \log{q} \leq 0$, $-1 \leq \log{s} \leq 1$, and $0\leq \alpha_0 \leq 2\pi$, respectively. We then perform MCMC analysis on the grid search result to identify the global minimum as well as the uncertainties on all parameters. Out of the seven events that go through these additional steps, three have their model $\chi^2$ substantially reduced.

The linear approximation for the lens orbital motion is subject to additional constraint. To select physically plausible solutions, the ratio between the projected kinetic energy and the potential energy is introduced
\begin{equation} \label{eqn:beta}
\beta = \frac{\kappa M_\odot {\rm yr}^2}{8 \pi^2}\frac{\pi_{\rm E}}{\theta_{\rm E}}\gamma^2\left( \frac{s_0}{\pi_{\rm E}+\pi_{\rm S}/\theta_{\rm E}} \right)^3; \quad  \Vec{\gamma} \equiv \left( \frac{{\rm d}s/{\rm d}t}{s_0},\frac{{\rm d}\alpha}{{\rm d}t} \right) ,
\end{equation}
where $\pi_{\rm S} \equiv {\rm AU}/D_{\rm S}$ is the source parallax \citep{Dong:2009}. Binary systems that are physically bound should have $\beta$ in the range 0--1, with values at both ends disfavored \citep{Poleski:2014}. We will investigate the impact of adding the $\beta$ constraint on the best-fit solutions with linear approximation in Section~\ref{sec:result}.

\section{Results} \label{sec:result}

\begin{table*} 
\centering
\caption{Detailed parameters of the 
60 simulated microlensing events, sorted in the order of increasing $\chi^2$ values. The events that are labelled as 1 to 40 in the first column has $q=1$.The events that are labelled as 41 to 60 has $q=0.3$. For each event, we include the randomly generated orbital parameters of the binary lens and the best-fit parameters (with 1-$\sigma$ uncertainties) of the microlensing model with linearized lens orbital motion. The minimum $\chi^2$ value of this fitting is also included, together with the parameter $\beta$ that quantifies the ratio of projected kinetic energy to the potential energy (Equation~\ref{eqn:beta}). The black line close to the end of the table separates the events that cannot be reasonably modeled with a linearized lens orbital motion.
}
\label{tab:result_input_param}
\begin{adjustbox}{scale=0.73}
\begin{threeparttable}
\begin{tabular}{c|c c c c| c c c c c c c c c c| c c}
\hline \hline 
 & \multicolumn{4}{c|}{Input Parameters}&\multicolumn{10}{c|}{Output Parameters}& & \\
\hline
Event & $\alpha_0$ & $l_0$ & $\omega$ & $I$ & $t_{0,\mathrm{fit}}$-$t_0$& $u_0$ & $t_{\rm E}$  & $\pi_{\rm E,N}$ & $\pi_{\rm E,E}$ & $q$ & $s_0 $& $\alpha_{0,\mathrm{fit}}$ & ${\rm d}s/{\rm d}t$ & ${\rm d}\alpha/{\rm d}t$  & $\chi^2$& $\beta$ \\
\# & Deg & Deg  &Deg  & Deg & Days & -& Days & - &-&-& $r_{\rm E}$& Deg&$r_{\rm E}$/yr & Deg/yr & -& - \\
\hline
38    & 96.130              & 354.354 & 295.781 & 86.005 & 0.0(4)         & 0.301(7)   & 100(3)    & 0.018(15)     & 0.021(20)     & 1.00(5)   & 0.801(8)   & 96.2(7)       & 0.29(3)          & 6(3)                    & 0.009                       & 0.163 \\
2     & 254.765             & 239.322 & 57.770  & 10.961 & 0.1(6)         & 0.301(9)   & 100.1(17) & 0.023(22)     & 0.021(17)     & 0.99(5)   & 0.801(9)   & 254.7(10)     & -0.11(4)         & 33(5)                   & 0.012                       & 0.454 \\
4     & 87.050              & 69.981  & 94.643  & 34.187 & -0.1(6)        & 0.299(14)  & 100(5)    & 0.014(19)     & 0.020(25)     & 1.00(8)   & 0.799(13)  & 87.2(9)       & 0.11(4)          & 31(4)                   & 0.058                       & 0.302 \\
50    & 353.098             & 350.788 & 310.805 & 89.820 & 0.1(5)         & 0.310(19)  & 98(4)     & 0.03(4)       & 0.016(18)     & 0.305(11) & 0.804(8)   & 353.2(7)      & 0.37(11)         & -3(7)                   & 0.062                       & 0.272 \\
24    & 269.33              & 315.975 & 175.277 & 66.434 & -0.1(3)        & 0.300(5)   & 99.1(21)  & 0.022(17)     & 0.017(16)     & 1.01(3)   & 0.800(6)   & 269.5(7)      & 0.081(8)         & -24(3)                  & 0.349                       & 0.216 \\
25    & 353.112             & 272.466 & 317.996 & 89.601 & -0.0(3)        & 0.310(18)  & 98(4)     & 0.01(3)       & 0.016(16)     & 1.01(3)   & 0.807(13)  & 353.1(7)      & -0.28(8)         & -4(7)                   & 0.440                       & 0.102 \\
15    & 49.853              & 31.426  & 326.806 & 29.142 & 0.1(8)         & 0.303(7)   & 99(3)     & 0.02(3)       & 0.017(20)     & 1.00(3)   & 0.804(8)   & 49.7(8)       & 0.04(6)          & -37(7)                  & 0.550                       & 0.398 \\
19    & 308.969             & 246.512 & 275.139 & 48.084 & -0.1(7)        & 0.298(9)   & 101(5)    & 0.03(4)       & 0.03(3)       & 1.00(3)   & 0.798(13)  & 309.1(7)      & 0.03(11)         & -27(9)                  & 0.579                       & 0.296 \\
46    & 40.033              & 336.113 & 155.950 & 45.451 & -0.1(10)       & 0.298(24)  & 100(7)    & 0.03(4)       & 0.02(4)       & 0.299(15) & 0.799(14)  & 40.0(17)      & 0.04(16)         & -30(22)                 & 0.597                       & 0.345 \\
54    & 31.766              & 73.830  & 152.136 & 69.800 & 0.2(7)         & 0.294(16)  & 102(4)    & 0.032(23)     & 0.025(16)     & 0.300(12) & 0.796(9)   & 31.3(11)      & -0.05(16)        & 20(18)                  & 0.630                       & 0.166 \\
21    & 269.347             & 192.574 & 80.161  & 15.797 & -0.1(6)        & 0.301(19)  & 100(9)    & 0.019(21)     & 0.02(3)       & 0.99(11)  & 0.800(14)  & 269.5(9)      & -0.01(4)         & 32(8)                   & 0.641                       & 0.307 \\
52    & 164.608             & 182.495 & 130.792 & 81.884 & 0.1(5)         & 0.31(4)    & 98(10)    & 0.02(4)       & 0.018(29)     & 0.305(23) & 0.805(16)  & 164.3(10)     & 0.24(17)         & 1(18)                   & 0.675                       & 0.114 \\
60    & 246.628             & 194.656 & 142.229 & 36.433 & 0.2(4)         & 0.303(11)  & 100.1(22) & 0.022(28)     & 0.023(15)     & 0.295(19) & 0.805(14)  & 246.6(6)      & 0.04(4)          & 27(6)                   & 0.700                       & 0.301 \\
33    & 226.227             & 108.501 & 16.311  & 71.487 & 0.3(7)         & 0.296(7)   & 101(3)    & 0.02(4)       & 0.023(16)     & 1.00(3)   & 0.797(8)   & 226.0(8)      & 0.29(7)          & -10(6)                  & 0.719                       & 0.212 \\
48    & 156.281             & 202.447 & 279.590 & 58.978 & -0.1(5)        & 0.296(12)  & 101(3)    & 0.02(3)       & 0.018(17)     & 0.299(9)  & 0.798(6)   & 156.5(9)      & 0.13(15)         & -20(10)                 & 0.721                       & 0.146 \\
55    & 263.063             & 265.132 & 175.231 & 68.898 & -0.1(5)        & 0.301(27)  & 98(10)    & 0.016(19)     & 0.011(29)     & 0.30(3)   & 0.80(4)    & 263.2(6)      & -0.31(5)         & -20(21)                 & 0.753                       & 0.180 \\
27    & 61.391              & 78.254  & 304.558 & 5.906  & 0.1(3)         & 0.300(4)   & 99.9(13)  & 0.030(16)     & 0.015(11)     & 0.996(18) & 0.801(6)   & 61.2(4)       & 0.14(4)          & -34.8(25)               & 0.776                       & 0.519 \\
22    & 175.303             & 66.647  & 69.709  & 59.578 & -0.1(3)        & 0.277(22)  & 104(7)    & 0.01(3)       & 0.036(19)     & 0.99(4)   & 0.783(21)  & 175.3(6)      & 0.20(6)          & -13(9)                  & 0.782                       & 0.148 \\
6     & 58.397              & 10.301  & 146.133 & 35.941 & 0.1(5)         & 0.301(4)   & 99.8(17)  & 0.026(21)     & 0.015(13)     & 1.002(24) & 0.803(6)   & 58.3(5)       & 0.07(4)          & -35(4)                  & 0.797                       & 0.474 \\
51    & 146.358             & 21.552  & 307.137 & 82.786 & 0.1(8)         & 0.273(24)  & 108(7)    & 0.00(3)       & 0.036(19)     & 0.284(18) & 0.785(13)  & 146.4(13)     & 0.09(19)         & 22(17)                  & 0.820                       & 0.196 \\
23    & 320.697             & 355.496 & 272.427 & 74.927 & 0.4(9)         & 0.307(11)  & 97(4)     & 0.03(4)       & 0.010(21)     & 1.01(3)   & 0.807(11)  & 320.3(9)      & -0.07(9)         & 24(7)                   & 0.835                       & 0.244 \\
53    & 181.890             & 236.335 & 151.392 & 29.038 & 0.0(4)         & 0.335(20)  & 93(3)     & 0.04(3)       & 0.000(15)     & 0.319(12) & 0.812(6)   & 181.4(6)      & -0.14(8)         & -37(5)                  & 0.836                       & 0.635 \\
1     & 204.651             & 341.812 & 7.493   & 61.817 & -0.1(8)        & 0.308(18)  & 98(4)     & 0.02(3)       & 0.012(22)     & 1.01(3)   & 0.806(13)  & 204.7(8)      & 0.10(9)          & 19(6)                   & 0.860                       & 0.103 \\
12    & 36.627              & 188.001 & 138.588 & 56.377 & 0.0(9)         & 0.297(12)  & 101(3)    & 0.02(4)       & 0.028(16)     & 0.98(3)   & 0.797(10)  & 36.7(8)       & 0.17(7)          & 20(7)                   & 0.879                       & 0.220 \\
58    & 138.967             & 204.001 & 264.957 & 38.056 & 0.1(8)         & 0.294(21)  & 102(6)    & 0.02(3)       & 0.029(25)     & 0.295(16) & 0.797(13)  & 138.7(13)     & -0.03(18)        & -27(23)                 & 0.959                       & 0.301 \\
36    & 229.239             & 328.150 & 201.076 & 87.025 & 0.3(3)         & 0.297(4)   & 101.0(11) & 0.03(3)       & 0.027(9)      & 0.997(17) & 0.798(4)   & 229.0(3)      & 0.22(3)          & -2(4)                   & 1.082                       & 0.100 \\
31    & 78.543              & 296.606 & 309.138 & 81.111 & -0.0(5)        & 0.300(8)   & 100.3(25) & 0.019(16)     & 0.016(15)     & 1.01(5)   & 0.798(8)   & 78.6(6)       & -0.39(3)         & 6(3)                    & 1.116                       & 0.247 \\
49    & 72.552              & 206.079 & 149.546 & 72.144 & 0.13(27)       & 0.299(5)   & 101.2(14) & 0.009(19)     & 0.025(10)     & 0.304(8)  & 0.7971(29) & 72.3(5)       & 0.076(22)        & 7(3)                    & 1.127                       & 0.028 \\
13    & 14.018              & 54.434  & 249.775 & 19.536 & -0.3(7)        & 0.337(25)  & 95(4)     & 0.03(3)       & 0.014(21)     & 0.95(4)   & 0.820(11)  & 13.9(8)       & 0.22(9)          & 26(6)                   & 1.311                       & 0.403 \\
37    & 124.091             & 145.739 & 270.582 & 88.978 & -0.1(4)        & 0.301(3)   & 99.7(7)   & 0.026(25)     & 0.017(10)     & 1.005(20) & 0.800(3)   & 124.1(4)      & -0.232(21)       & -1.3(23)                & 1.316                       & 0.104 \\
57    & 224.995             & 150.120 & 208.660 & 40.942 & 0.3(3)         & 0.2986(24) & 101.0(11) & 0.018(17)     & 0.025(9)      & 0.296(3)  & 0.7996(27) & 224.77(25)    & 0.019(22)        & -25(3)                  & 1.349                       & 0.232 \\
35    & 201.746             & 308.222 & 280.904 & 74.102 & 0.2(5)         & 0.298(11)  & 100(3)    & 0.03(3)       & 0.023(16)     & 1.00(3)   & 0.799(9)   & 201.5(6)      & -0.31(7)         & 11(6)                   & 1.381                       & 0.244 \\
45    & 157.899             & 315.834 & 334.946 & 86.212 & -0.1(7)        & 0.31(3)    & 99(8)     & 0.01(4)       & 0.026(26)     & 0.301(24) & 0.802(15)  & 157.9(13)     & -0.34(20)        & -6(24)                  & 1.393                       & 0.228 \\
7     & 308.163             & 287.994 & 161.938 & 4.198  & 0.4(5)         & 0.299(5)   & 98(3)     & 0.02(3)       & 0.005(19)     & 1.012(25) & 0.799(7)   & 307.8(6)      & -0.16(6)         & -28(8)                  & 1.471                       & 0.177 \\
47    & 268.402             & 219.577 & 307.052 & 85.130 & -0.1(4)        & 0.312(24)  & 95(10)    & 0.035(17)     & 0.003(25)     & 0.289(28) & 0.82(3)    & 268.6(7)      & 0.15(5)          & -13(22)                 & 1.471                       & 0.115 \\
28    & 168.055             & 333.483 & 45.141  & 50.856 & 0.1(4)         & 0.326(9)   & 96.7(22)  & 0.01(3)       & 0.020(16)     & 0.97(3)   & 0.814(6)   & 168.3(6)      & -0.15(8)         & 19(3)                   & 1.483                       & 0.155 \\
41    & 12.994              & 94.908  & 63.725  & 67.816 & -0.2(6)        & 0.29(4)    & 102(8)    & -0.01(4)      & 0.03(3)       & 0.296(23) & 0.797(13)  & 13.3(10)      & 0.05(22)         & 21(19)                  & 1.541                       & 0.165 \\
59    & 338.357             & 116.179 & 315.211 & 43.467 & -0.1(5)        & 0.279(13)  & 106(4)    & 0.01(4)       & 0.040(15)     & 0.285(10) & 0.789(8)   & 338.4(6)      & 0.01(7)          & -24(4)                  & 1.701                       & 0.230 \\
26    & 185.745             & 344.559 & 305.402 & 0.964  & -0.3(5)        & 0.38(3)    & 89.0(24)  & 0.03(3)       & 0.003(16)     & 1.03(4)   & 0.841(7)   & 185.7(7)      & -0.07(11)        & 22(6)                   & 1.786                       & 0.217 \\
40    & 315.663             & 300.004 & 121.919 & 81.167 & 0.2(6)         & 0.303(8)   & 98(3)     & 0.03(4)       & 0.004(22)     & 1.01(3)   & 0.805(9)   & 315.4(6)      & -0.30(8)         & -2(5)                   & 1.821                       & 0.180 \\
44    & 332.541             & 141.592 & 320.053 & 50.501 & -0.2(7)        & 0.289(7)   & 104(3)    & -0.01(5)      & 0.041(18)     & 0.292(7)  & 0.791(6)   & 332.9(8)      & 0.08(6)          & -25(4)                  & 1.914                       & 0.264 \\
39    & 82.302              & 208.853 & 236.866 & 75.330 & -0.07(22)      & 0.294(4)   & 98.2(14)  & 0.027(8)      & 0.010(11)     & 0.96(3)   & 0.795(5)   & 82.2(4)       & -0.199(13)       & -11(2)                  & 1.950                       & 0.118 \\
56    & 193.935             & 103.689 & 8.906   & 60.962 & 0.7(5)         & 0.245(14)  & 115(5)    & 0.00(3)       & 0.064(15)     & 0.268(10) & 0.774(9)   & 194.2(6)      & 0.41(5)          & -5(4)                   & 2.260                       & 0.242 \\
30    & 209.084             & 257.353 & 339.898 & 6.941  & -0.3(5)        & 0.314(7)   & 97.4(24)  & -0.01(3)      & 0.012(15)     & 1.02(3)   & 0.811(7)   & 209.3(5)      & -0.19(8)         & 30(4)                   & 2.649                       & 0.117 \\
29    & 85.657              & 358.580 & 76.765  & 51.111 & -0.1(4)        & 0.293(9)   & 96.8(12)  & 0.020(20)     & 0.005(11)     & 0.96(7)   & 0.793(4)   & 85.7(8)       & -0.110(7)        & -31(8)                  & 2.924                       & 0.227 \\
5     & 132.001             & 91.733  & 195.571 & 42.250 & 0.4(9)         & 0.285(8)   & 106(5)    & 0.09(4)       & 0.024(22)     & 0.98(3)   & 0.782(11)  & 131.4(7)      & 0.11(7)          & 25(8)                   & 3.071                       & 0.104 \\
9     & 126.198             & 92.594  & 299.857 & 23.085 & 0.6(5)         & 0.304(5)   & 95.8(22)  & 0.03(3)       & -0.005(18)    & 0.999(23) & 0.803(7)   & 125.8(5)      & 0.13(5)          & -25(6)                  & 3.141                       & 0.244 \\
3     & 233.906             & 124.058 & 96.952  & 41.645 & 0.2(5)         & 0.300(5)   & 100.2(12) & 0.04(3)       & 0.019(9)      & 0.979(18) & 0.800(7)   & 233.7(5)      & -0.02(6)         & 25(4)                   & 3.401                       & 0.252 \\
11    & 163.712             & 115.708 & 54.244  & 78.076 & 0.3(6)         & 0.287(23)  & 104(4)    & 0.02(4)       & 0.028(20)     & 0.96(4)   & 0.788(14)  & 163.2(7)      & -0.04(10)        & 16(6)                   & 6.139                       & 0.104 \\
43    & 83.448              & 162.720 & 167.129 & 57.535 & 0.56(18)       & 0.3143(18) & 101.1(6)  & 0.033(14)     & 0.032(5)      & 0.338(4)  & 0.8036(21) & 82.2(3)       & 0.155(6)         & 3.4(13)                 & 7.211                       & 0.053 \\
32    & 77.802              & 198.425 & 144.119 & 89.662 & 0.50(22)       & 0.293(4)   & 101.7(10) & 0.033(5)      & 0.020(6)      & 0.992(22) & 0.793(5)   & 76.9(3)       & 0.206(9)         & 5.4(23)                 & 7.637                       & 0.100 \\
20    & 293.07              & 6.806   & 79.679  & 37.291 & 0.6(6)         & 0.291(9)   & 101.2(10) & 0.03(4)       & 0.024(21)     & 1.03(6)   & 0.789(9)   & 292.2(7)      & -0.05(5)         & -36(7)                  & 8.701                       & 0.514 \\
16    & 118.781             & 298.482 & 94.457  & 6.209  & 2.9(6)         & 0.295(6)   & 95.0(18)  & 0.07(8)       & -0.03(3)      & 0.86(5)   & 0.789(14)  & 116.1(8)      & -0.07(4)         & 40(13)                  & 13.065                      & 0.387 \\
10    & 179.483             & 308.066 & 300.682 & 8.894  & 0.3(6)         & 0.414(18)  & 85.7(19)  & 0.04(4)       & -0.011(17)    & 1.07(4)   & 0.852(5)   & 179.5(8)      & -0.22(11)        & 6(4)                    & 13.669                      & 0.126 \\
42    & 49.839              & 211.637 & 6.609   & 82.064 & -3.1(14)       & 0.282(5)   & 104.0(20) & -0.009(24)    & 0.027(16)     & 0.239(19) & 0.787(9)   & 56.4(25)      & 0.02(18)         & 24(12)                  & 18.866                      & 0.200 \\
14    & 215.925             & 60.330  & 39.943  & 88.510 & -0.3(5)        & 0.301(5)   & 96.1(18)  & 0.06(3)       & -0.001(13)    & 1.029(21) & 0.810(5)   & 216.4(6)      & 0.24(5)          & 18(6)                   & 40.764                      & 0.217 \\
\hline
34    & 118.770             & 203.967 & 5.335   & 81.054 & -0.5(3)        & 0.3062(23) & 100.6(7)  & 0.002(17)     & 0.039(8)      & 1.014(15) & 0.808(5)   & 119.1(3)      & -0.13(3)         & 0.8(18)                 & 172.813                     & 0.038 \\
8     & 245.904             & 246.009 & 223.787 & 69.655 & -1.0(3)        & 0.294(3)   & 100.5(12) & 0.009(15)     & 0.039(10)     & 1.133(21) & 0.808(5)   & 247.5(4)      & -0.16(4)         & -11(3)                  & 215.648                     & 0.108 \\
18    & 239.032             & 303.966 & 85.041  & 81.645 & 2.1(5)         & 0.301(5)   & 101.9(15) & 0.04(3)       & 0.024(10)     & 0.909(14) & 0.802(7)   & 236.5(4)      & -0.14(5)         & 2(4)                    & 257.239                     & 0.040 \\
17    & 261.578             & 155.175 & 345.757 & 6.586  & 9.7(6)         & 0.293(15)  & 103.7(14) & 0.47(7)       & -0.144(25)    & 1.37(15)  & 0.930(9)   & 245.4(11)     & -0.31(9)         & 3(8)                    & 3994.755                    & 0.001 \\

\hline \hline
\end{tabular}
\end{threeparttable}
\end{adjustbox}
\end{table*}

\begin{figure}
\includegraphics[width=\columnwidth]{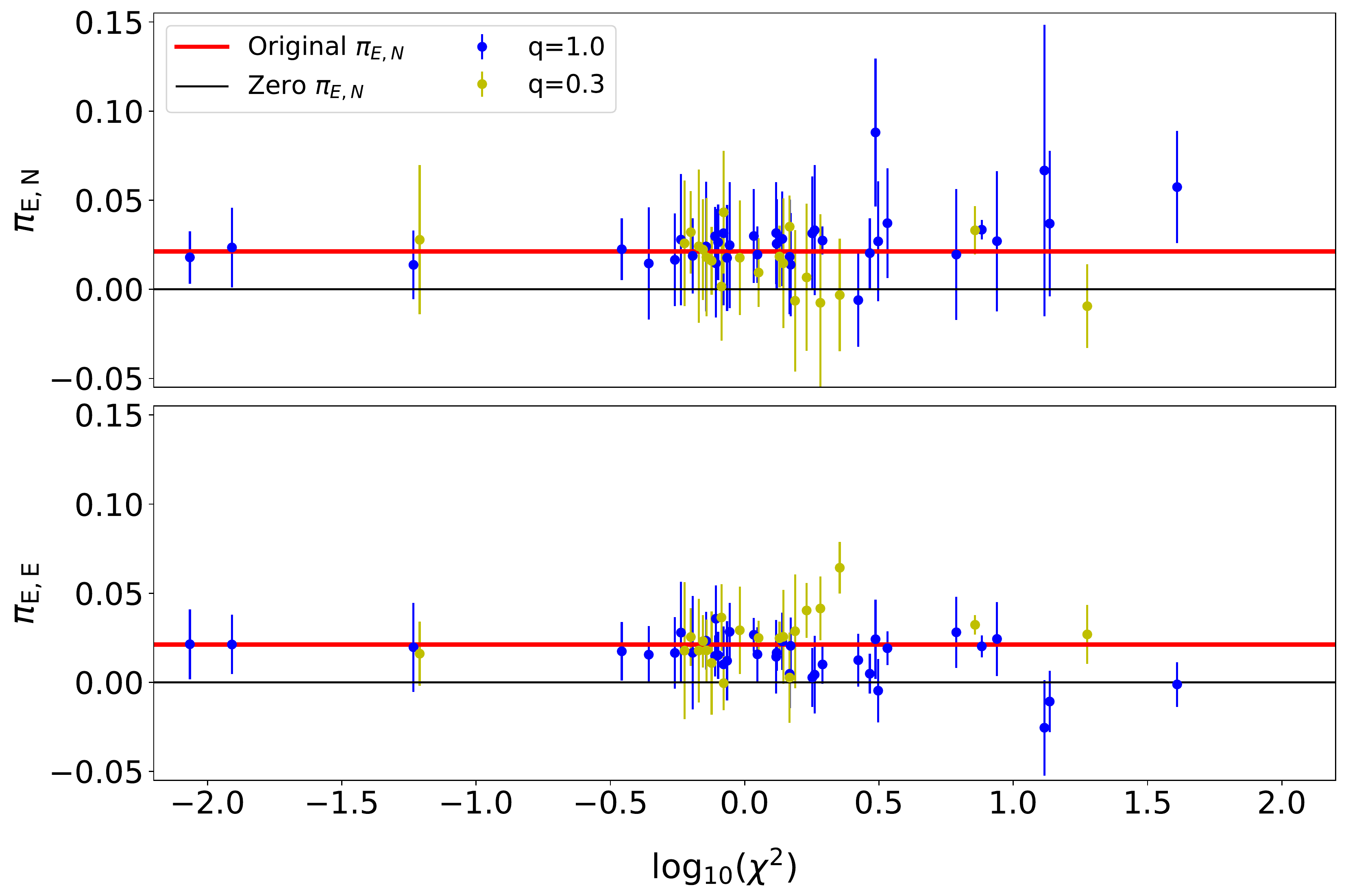}
\caption{The best-fit microlensing parallax parameters and the associated uncertainties as functions of the minimum model $\chi^2$ for simulated events that can be fit with the linearized lens orbital model. The blue dots indicate the events with $q=1$ and the yellow dots indicate the events with $q=0.3$.}
\label{fig:pi_scatter}
 \end{figure}

We provide the input parameters as well as the best-fit solution of all 60 simulated events in Table~\ref{tab:result_input_param}.

\begin{figure*}
\centering
\includegraphics[width=0.8\textwidth]{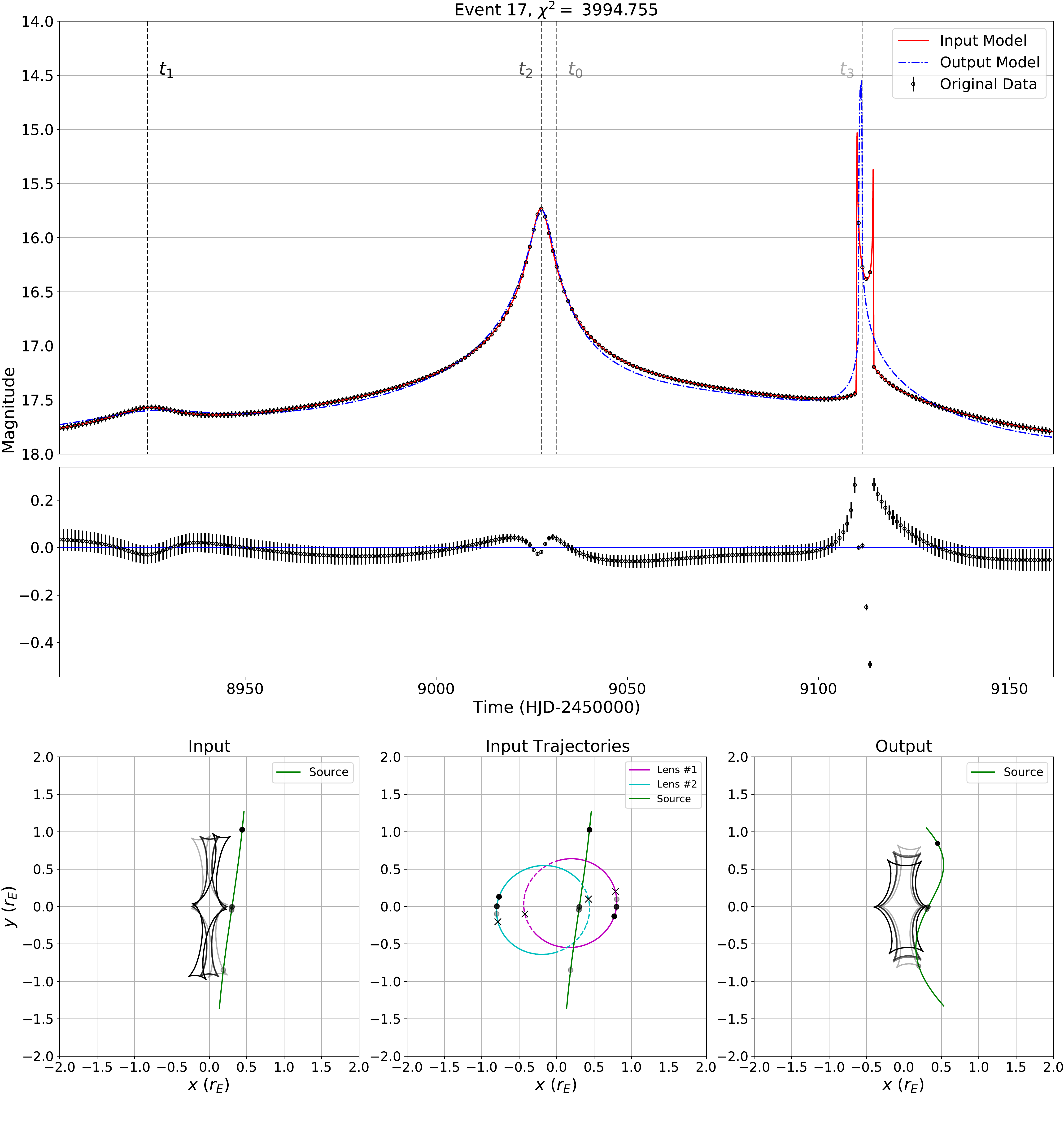}
\caption{One of the simulated events in which the linear approximation of the orbital motion fails. This event has the highest model $\chi^2$. The top panel shows the simulated light curve, the input model with full orbital motion, and the best-fit model with linear orbital motion. The middle panel shows the residuals of the fitting. The bottom panels are the lensing geometry of the input model (bottom left and bottom middle) and of the best-fit model with linear orbital motion (bottom right). 
Unlike the bottom left and bottom right panels that show the source trajectories and caustic structures, the bottom middle panel shows the source trajectory and the trajectories of the binary lenses.
The epoch at $t_0$ and three other epochs that correspond to noticeable binary features are indicated in the light curve. The source positions, lens positions, and caustic curves at these chosen epochs are also indicated in the bottom panels.}
\label{fig:example-lc-2}
\end{figure*}

\begin{figure}
\includegraphics[width=\columnwidth]{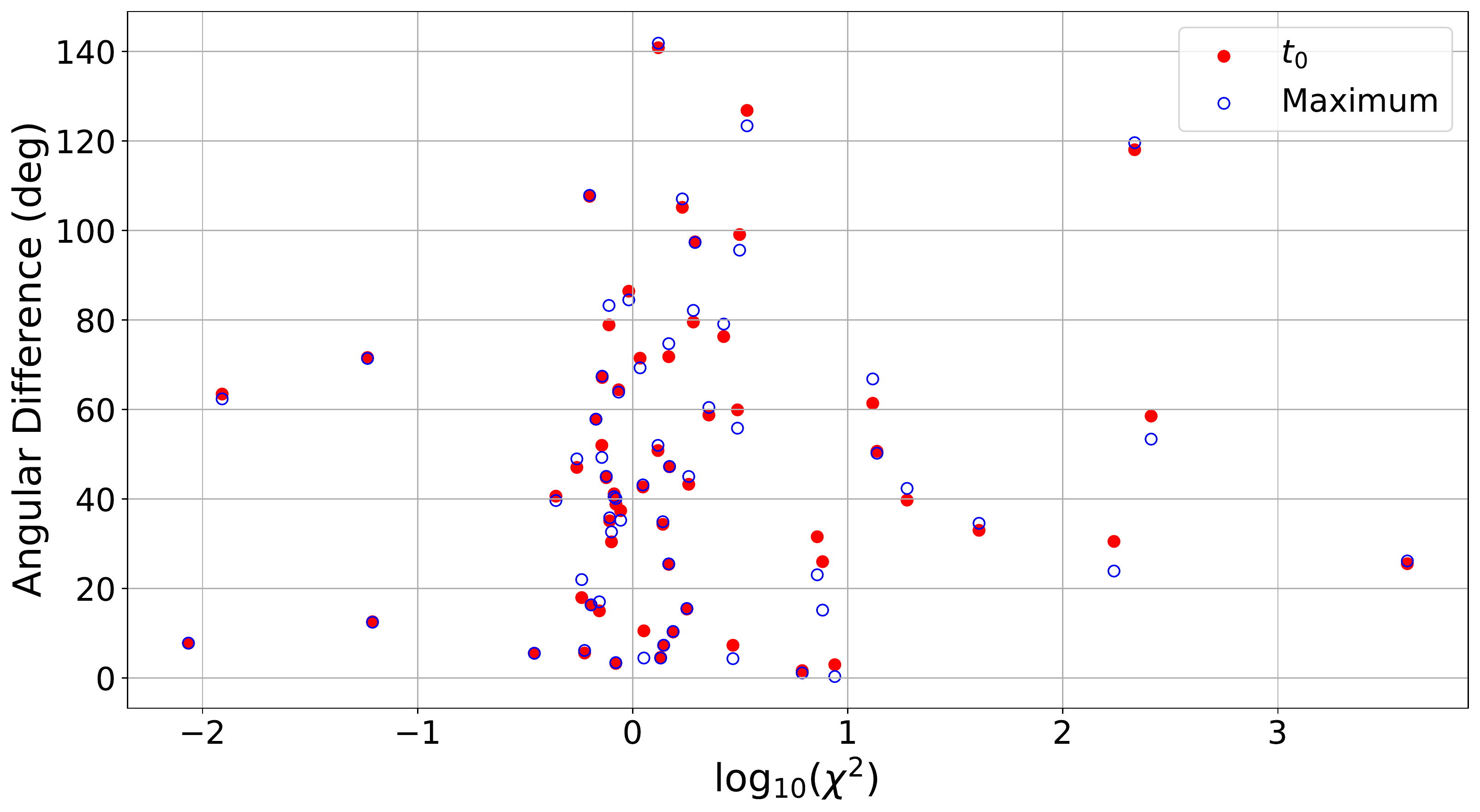}
\caption{Angular difference between the lens position at the chosen epoch ($t_0$ for the red filled circles and the epoch with the maximum magnification for the blue open dots) and the apocenter/pericenter of the ellipse-like lens trajectory. }
\label{fig:angle_diff}
\end{figure}

\begin{figure}
\includegraphics[width=\columnwidth]{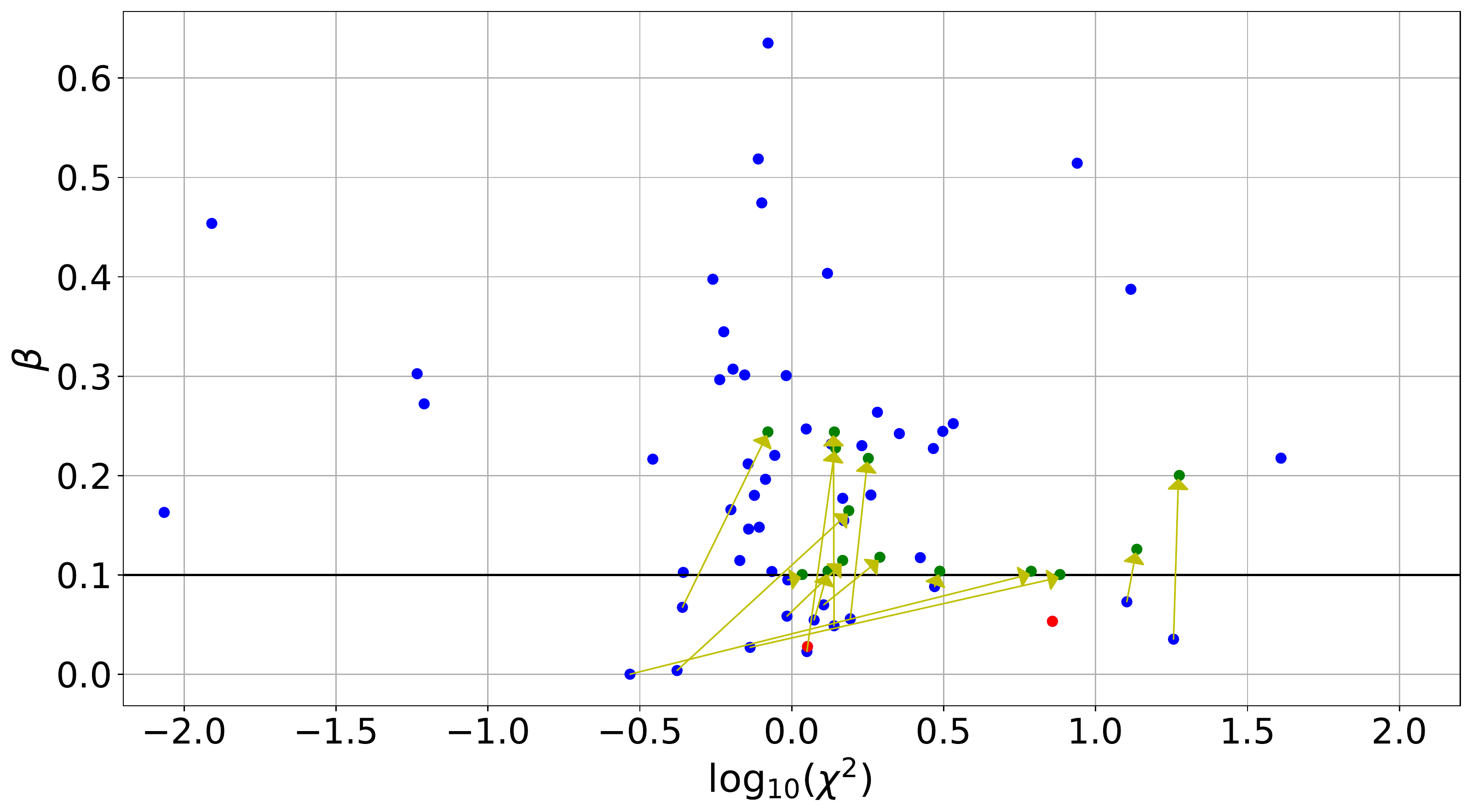}
\caption{The ratio between the projected kinetic and potential energies, $\beta$, for the well-fit (model minimum $\chi^2<100$) events in our simulation. The yellow lines connect the locations of the same event before and after an lower limit on $\beta$ ($>0.1$) is imposed. There are two events whose $\beta$ values are never above the chosen 0.1 limit. These events are indicated by the red dots.}
\label{fig:beta_scatter}
\end{figure}

The cumulative distribution of the best-fit $\chi^2$ values of all simulated events is shown in Figure~\ref{fig:chi2-cdf}. We adopt $\chi^2=100$ as the threshold, below which the fit is considered reasonable. This seemingly large threshold is to take into account the systematic noise that is not included in our simulation but exists in real observations. Such systematics can usually affect the goodness-of-fit by $\Delta\chi^2$ of dozens. Furthermore, our result is largely insensitive to the choice of $\chi^2$ threshold. As shown in Figure~\ref{fig:chi2-cdf}, the change of this $\chi^2$ threshold from 100 to 10 will only affect four events.

As shown in Figure~\ref{fig:chi2-cdf}, all except four of the simulated events with full orbital motions can be reasonably modeled with the linear orbital motion approximation. For the well-fit events, we show in Figure~\ref{fig:pi_scatter} the uncertainties on the parallax parameters, $\pi_{\rm E,E}$ and $\pi_{\rm E,N}$, derived from the MCMC analyses. Overall, the uncertainties on the parallax parameters are substantial compared to the amplitude of the microlensing parallax, rendering no (secure) detection of the microlensing parallax effect in any of the well-fit events. This agrees with the previous study on detecting isolated BHs that the microlensing parallax is generally too small to be robustly detected \citep{Karolinski_2020}. The null detection of microlensing parallax effect suggests that the mass of the lens system cannot be well determined. This provides a plausible explanation for the null detection of BH binaries via microlensing. It is also worth pointing out that none of the well-fit events has secure parallax detection with large amplitude (e.g., $\pi_{\rm E}\gtrsim0.1$), suggesting that, at least for well sampled BH binary events, the degeneracy between binary orbital motion and microlensing parallax is rare.

Out of the 60 simulated events with full orbital motion, four cannot be modeled with the linear orbital motion approximation. We show in Figure~\ref{fig:example-lc-2} the one with the highest $\chi^2$. The remaining three can be found in the online-only figures (Figures 7--9). The simulated light curves for 56 well-fit events are also shown in online-only figures (Figure 10).
Why is the linear orbital motion assumption broken in these cases? Perhaps the most straightforward hypothesis is that at the peak of the event the lens is at (or close to) the peri- or apo-center of the ellipse-shaped trajectory (see the bottom middle panel of Figure~\ref{fig:example-lc-2}), where the time derivative of the projected separation  $s$ is about to switch signs and thus the linear approximation of $s$ is no longer valid. To test this hypothesis, we have calculated the angular distances of the lens position at $t_0$ to the pericenter (or apocenter, whichever is closer) of the ellipse-like trajectory for all simulated events. The results are shown in Figure~\ref{fig:angle_diff} as the red dots. Although events that cannot be well fit tend to have relatively small angular separations and their lenses are preferentially closer to the pericenter/apocenter, a substantial fraction of the well-fit events have similarly small (or even smaller) angular separations. This general trend holds even when the time of maximum magnification in the simulated data is used for computing the angular distances. Therefore, passing through the pericenter/apocenter of the ellipse-shaped lens trajectory during the course of the microlensing event is probably not the primary (or the only) reason for the invalidity of the linear approximation of the lens orbital motion. The reason why the linear orbital motion assumption is broken for some events remains unclear. Further investigations are needed.

When the linear approximation of the lens orbital motion is used, the parameter $\beta$ (Equation~\ref{eqn:beta}) is usually employed to evaluate the validity of the orbital motion \citep[e.g.,][]{Dong:2009}. 
As listed in Table~\ref{tab:result_input_param} and illustrated in Figure~\ref{fig:beta_scatter}, the events that can be modeled with the linear orbital motion approximation all have reasonably small values of $\beta$. Considering that the extremely small values are also disfavored \textit{a priori} \citep{Poleski:2014}, we impose a lower limit on the parameter $\beta$ at 0.1 on the MCMC results for well-fit events and find that the goodness of fit hardly changes, expect for two events that have all MCMC samples with $\beta<0.1$ This suggests that the events that can be modeled with linear orbital motion are fairly robust against the variation of $\beta$.

\section{Conclusion and Discussion} \label{sec:discussion}

While the number of expected binary BH detections via microlensing is highly uncertain, it is true that microlensing is capable of detecting such systems and yet zero has been unambiguously identified so far. A direct determination of the lens mass remains the key to distinguish the BH systems from the common stellar binaries. This work studies the detectability of the microlensing parallax effect, one of the two ingredients into a direct mass determination.

We simulated 60 microlensing events that are assumed to be produced by typical binary BHs with full orbital motion. The simulated events were then modeled in the standard approach used in current microlensing analysis. In particular, the model assumes linear orbital motion of the binary lens system, based on the fact that the duration of the microlensing event is typically much shorter than the orbital period of the binary. We report a few new findings:
\begin{itemize}
    \item Even though the microlensing events caused by binary BH lenses have preferentially long timescales, the microlensing parallax effect in typical binary BH events cannot be reliably detected because of the small amplitude of the microlensing parameter. Given the crucial role of the parallax parameter in determining the mass of dark microlenses, the null detection of the parallax effect may be partially responsible for the null detection of binary BHs from microlensing.
    \item The linear orbital motion approximation that is commonly used in microlensing modelings may fail in a small fraction ($\lesssim 7\%$) of binary events. This is generally consistent with that is practiced in the analyses of real events. Wherever the linear approximation fails, the full orbital motion must be taken into account \citep[e.g.,][]{Skowron:2011}. A robust detection of the full orbital motion can in principle provide another relation between the lens mass and lens distance, which, once combined with the angular Einstein radius measurement from caustic crossings \citep{Yoo:2004}, may also lead to a direct determination of the lens properties.
    \item Lens orbital motion does not usually produce erroneous microlensing parallax, at least in typical BH events. This is probably because the lens orbital motion has more flexibility than the microlensing parallax effect.
\end{itemize}

Our simulations were performed with selected choices of microlensing parameters. These values were chosen as typical for microlensing events with BH binary lenses based on the population synthesis simulation of \citet{Lam:2020}, as observations have yielded very limited constraints on the distribution of Galactic BH binaries with wide separations. Additionally, our simulated observations are rather ideal, with no seasonal gaps or blending effect and relatively bright sources of fixed magnitude. The inclusion of these realistic conditions will further reduce the detectability of the microlensing parallax and thus strengthen the conclusion. We expect that the results reported here are qualitatively applicable to reality and provide useful guidance toward the search for Galactic BH binaries.

In the present work, the way that the microlensing parallax effect is to be detected is through the annual oscillation of Earth motion around Sun. An alternative way that can also measure the microlensing parallax is through simultaneous observations from at least two well-separated observatories \citep{Refsdal:1966, Gould:1994}. The recent practice employing the \emph{Spitzer} and two-wheeled \emph{Kepler} (\emph{K2}) space telescopes has yielded the microlensing parallax detections of $\sim1000$ microlensing events \citep[e.g.,][]{Dong:2007, Udalski:2015b, Henderson:2016, Zhu:2017}. This satellite parallax method has also been proposed for future space-based microlensing surveys,
including \emph{Roman} microlensing survey \citep{Penny:2019} and the proposed microlensing surveys with \emph{Euclid} \citep{Penny:2013} and the Chinese Space Station Telescope \citep{Yan:2022}. However, this method also encounters difficulties in measuring/constraining parallax parameter, $\pi_{\rm E}$, to the precision that is required for precise lens mass determinations. The follow-up mode and the relatively short observing window of the \emph{Spitzer} microlensing campaign imply that the satellite usually could not precisely constrain the microlensing parameters alone \citep{Yee:2015}. For the proposed future microlensing surveys at Earth--Sun L2 or low-Earth orbit, the spatial separation is not large enough to permit precise parallax measurements of long-timescale events. Therefore, it is plausible that our conclusion applies regardless of the detection method and that the microlensing parallax effect cannot be precisely measured/constrained in typical BH binary events.

Binary BH systems from the microlensing technique can complement the BH detections from other channels and thus provide an unbiased picture of the BH demographics. Our work points out a few characteristic features of binary BH events that have been overlooked in previous searches. The results reported here may therefore help to more efficiently identify the rare but interesting BH events from the dominating, normal microlensing events.

\section*{Acknowledgements}

We would like to thank Subo Dong, Valerio Bozza, and an anonymous referee for comments and suggestions.
We acknowledge the science research grants from the China Manned Space Project with No.\ CMS-CSST-2021-B12. This work is supported by the National Science Foundation of China (grant No. 12173021 and 12133005).
We would also like to thank Weicheng Zang for assistance with the binary-lens modeling.

\section*{Data Availability}

The data underlying this article will be shared on reasonable request to the corresponding author.






\setcounter{figure}{6}

\thispagestyle{empty}
\begin{figure*}
\centering
\includegraphics[width=0.98\textwidth]{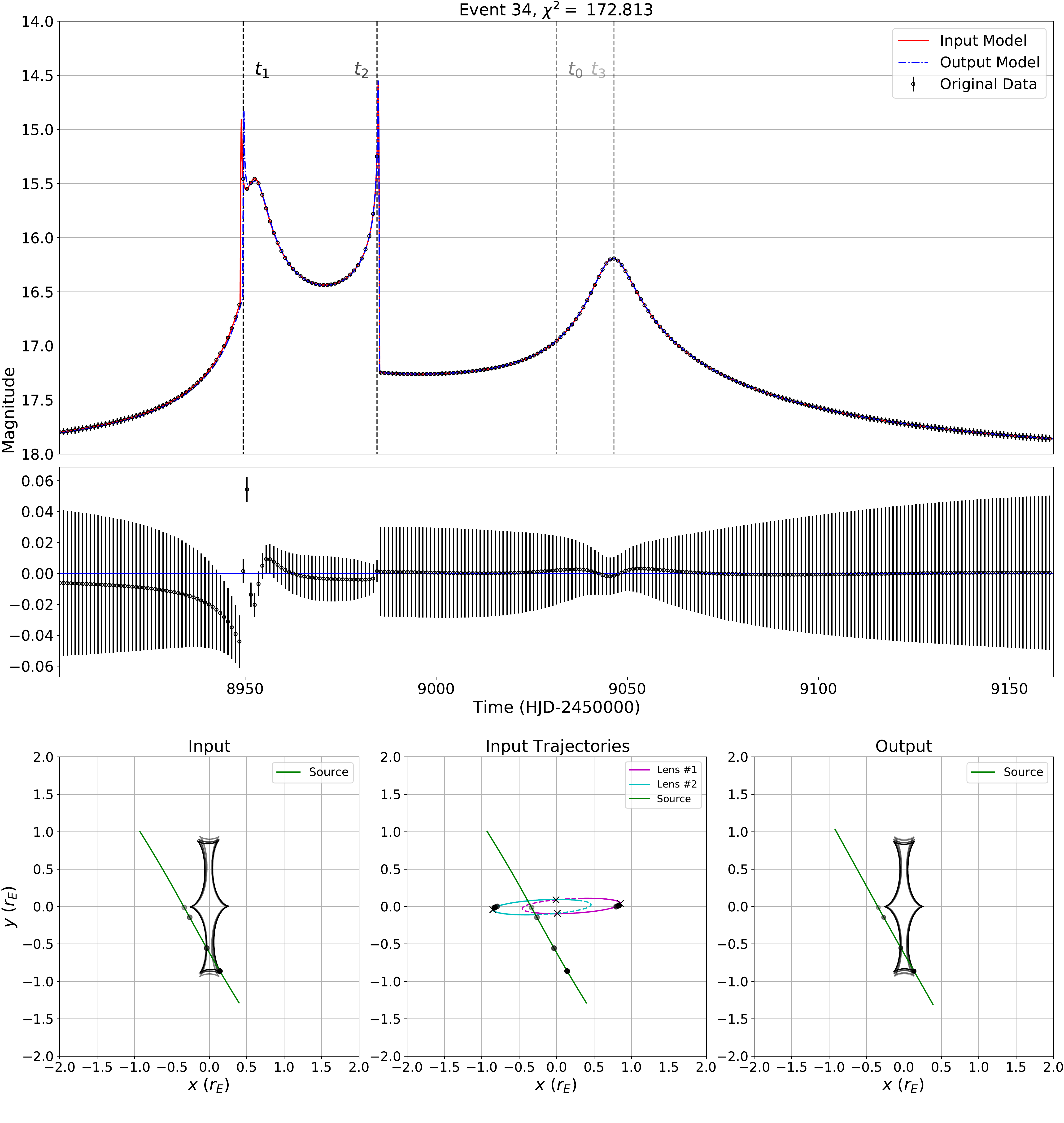}
\caption{The graph is similar to Figure 4 in the main paper, but for event 34 with $\chi^2$ = 172.813.}
\label{fig:example-lc-3}
\end{figure*}

\thispagestyle{empty}
\begin{figure*}
\centering
\includegraphics[width=0.98\textwidth]{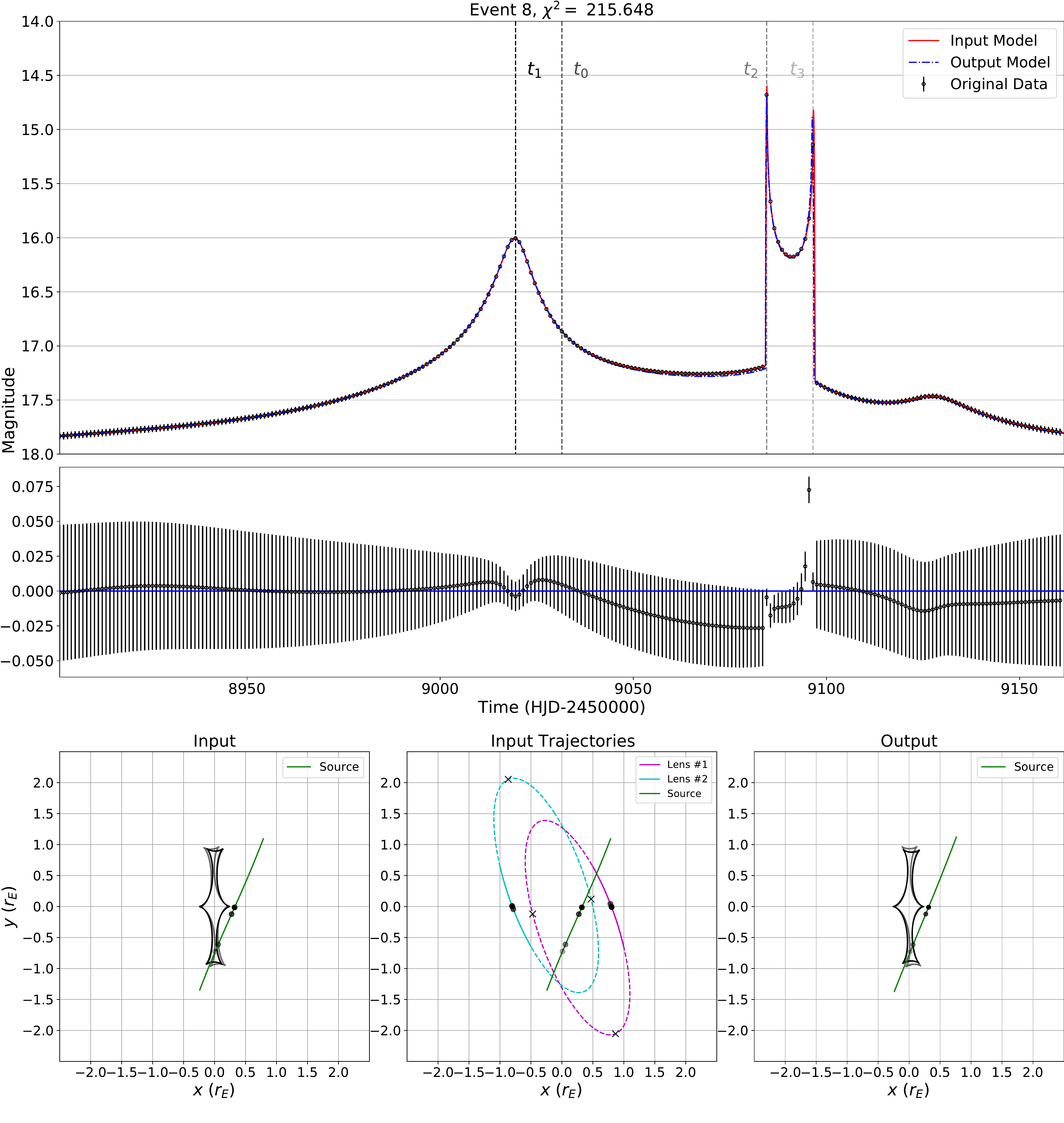}
\caption{The graph is similar to Figure 4 in the main paper, but for event 8 with $\chi^2$ = 215.648.}
\label{fig:example-lc-4}
\end{figure*}

\thispagestyle{empty}
\begin{figure*}
\centering
\includegraphics[width=0.98\textwidth]{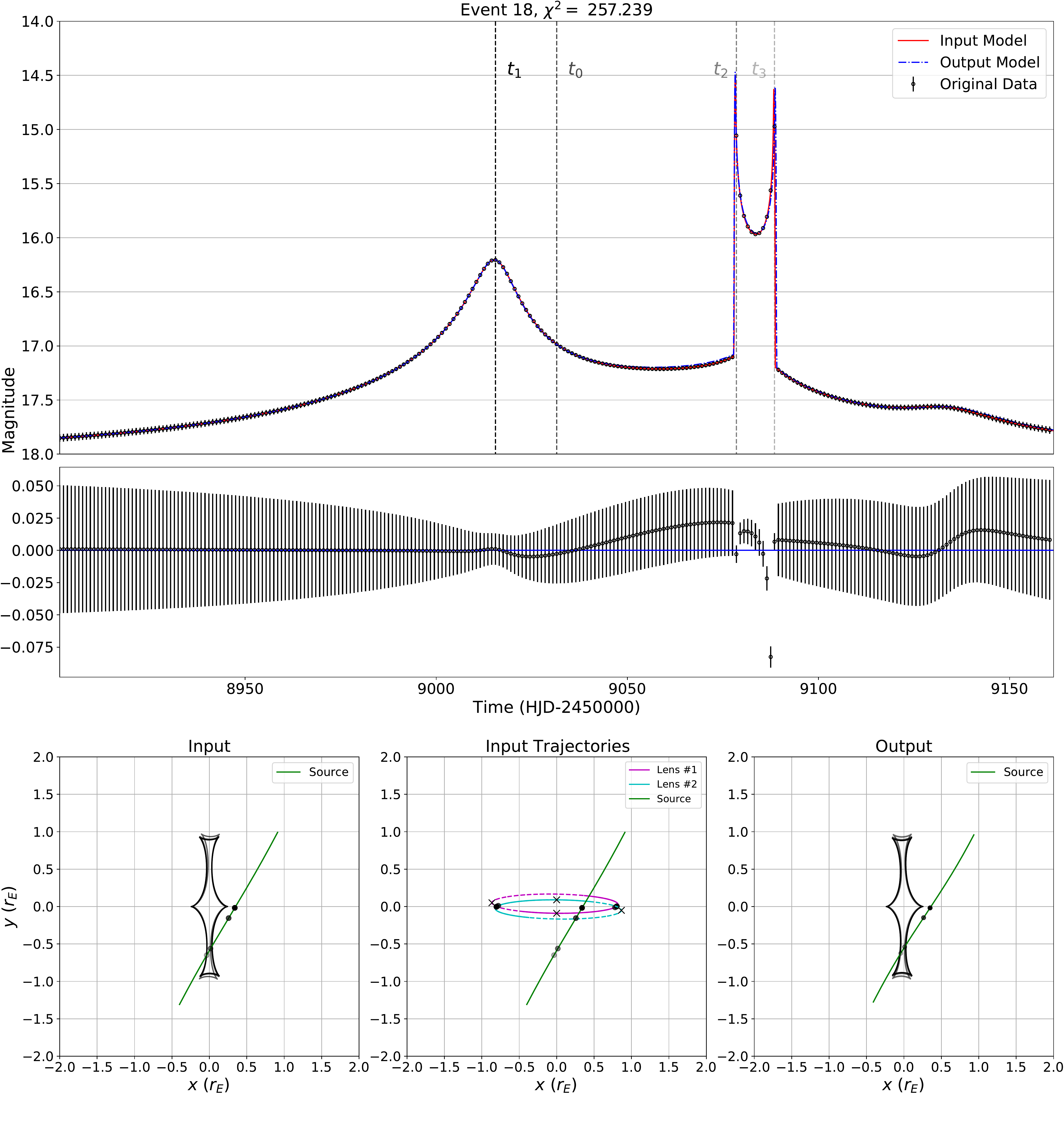}
\caption{The graph is similar to Figure 4 in the main paper, but for event 18 with $\chi^2$ = 257.239.}
\label{fig:example-lc-5}
\end{figure*}

\thispagestyle{empty}
\begin{figure*}
\centering
\includegraphics[width=0.98\textwidth]{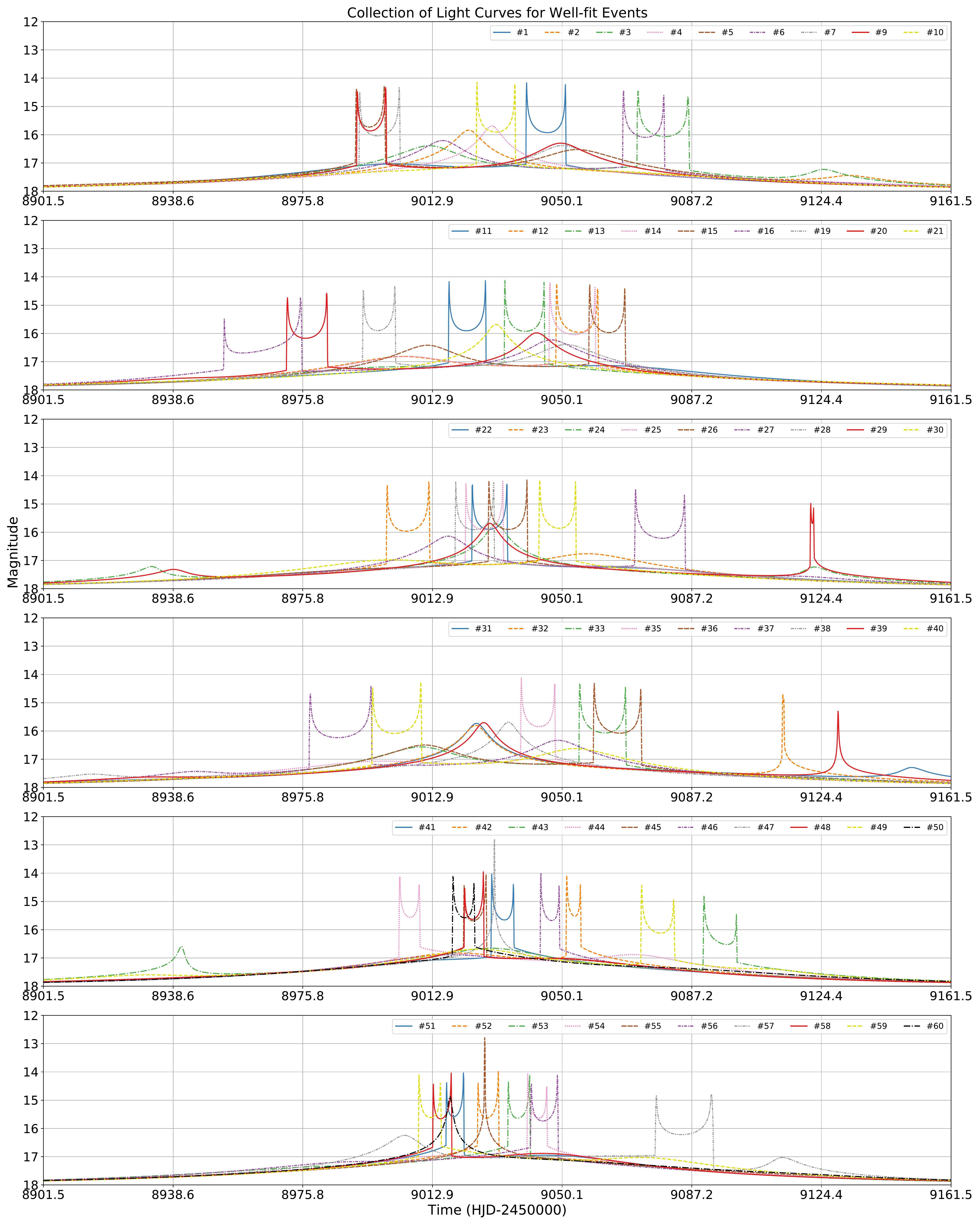}
\caption{The graph is a collection of simulated light curves for 56 well-fit events within 260-day window centered at $t_0$.}
\label{fig:example-lc-all}
\end{figure*}

\bibliographystyle{mnras}
\bibliography{ms1} 

\bsp	
\label{lastpage}
\end{CJK*}
\end{document}